[A

\def\vk{\bf k}

\def\bpm{\beta_{\pm}}
\def\bpl{\beta^+}
\def\bmn{\beta^-}
\def\ok{\omega_{\bf k}}
\def\a{\alpha}

\def\bp{{\beta_{ij}}}
\def\bm{{{\beta'}_{ij}}}
\def\bb{\bar\beta_{ij}}
\def\bpl{\beta_+}
\def\bmn{\beta_-}

\def\ha{{1\over 2}}

\def\be{\begin{equation}}
\def\te{\end{equation}}
\def\bea{\begin{eqnarray}}
\def\nn{\nonumber}
\def\tea{\end{eqnarray}}

\def\a{\alpha}

\def\l{\lambda}

\def\n{\nu}
\def\o{\omega}

\def\r{\rho}

\newskip\humongous \humongous=0pt plus 1000pt minus 1000pt

\newif\ifdtup

\documentstyle[12pt]{article}

\makeatletter                    
\@addtoreset{equation}{section}  
\makeatother                     

\begin{document}

\title{A Fluctuation-Dissipation Relation for Semiclassical Cosmology}

\author{B. L. Hu\thanks{ Email: hu@umdhep.umd.edu}\\
{\small Department of Physics, University of Maryland,
College Park, MD 20742, USA} \\
Sukanya Sinha\thanks{ Email: ssinha@iucaa.ernet.in}\\
{\small IUCAA, Post Bag 4, Ganeshkhind,  Pune 411007, India}}
\date{\today}
\maketitle
\centerline{(umdpp 93-164) }

\begin{abstract}
Using the concept of open systems where the classical geometry is treated as
the system and  the quantum matter field as the environment,
we derive a fluctuation-dissipation theorem for semiclassical cosmology.
This theorem which exists under very general conditions for dissipations
in the dynamics of the system, and the noise and fluctuations in the
environment, can be traced to the formal mathematical relation
between the dissipation and noise kernels
of the influence functional depicting the open system, and is ultimately a
consequence of the unitarity of the closed system. In particular,
for semiclassical gravity, it embodies the backreaction effect
of matter fields on the dynamics of spacetime. The backreaction equation
derivable from the influence action is in the form of a Einstein-Langevin
equation. It contains
a dissipative term in the equation of motion for the dynamics of spacetime
and a noise term related to the fluctuations of particle creation in the matter
field. Using the well-studied model of a quantum scalar field in a Bianchi
Type-I
universe we illustrate how this Langevin equation and the noise term are
derived
and show how the creation of particles and the dissipation of anisotropy
during the expansion of the universe can be understood as a manifestation
of this fluctuation-dissipation relation.
\end{abstract}

\newpage

\section{Introduction}

An important relation between the dissipation in the dynamics of a system
and the fluctuations in a heat bath with which the system interacts is the
fluctuation-dissipation relation (FDR) \cite{fdr}.
A first example of its manifestation is the Nyquist noise in an electric
circuit
\cite{Nyquist}. This relation is of practical interest in the design of noisy
systems \cite{noise}. It is also of theoretical interest in statistical physics
because it is a categorical relation which exists
between the  stochastic behavior of many microscopic particles
and the  deterministic behavior of a macroscopic system. It is therefore also
useful for the description of the interaction of a system with fields,
such as effects related to radiation reaction and vacuum fluctuations
between atoms and fields in quantum optics \cite{qo}.
The form of the FDR is usually given under near-equilibrium
conditions via linear response theory (LRT) \cite{lrt}.
We will see in this paper that this
relation has a much wider scope and a broader implication than has been
understood before. In particular we want to address problems involving gravity
and quantum fields in black holes and the early universe; and we are interested
in seeing this relation validated and implemented under non-equilibrium
conditions for quantum fields in curved spacetimes \cite{BirDav}.
The problem we choose for illustration is the backreaction of particles created
\cite{Par69,SexUrb,Zel,Hu74,Gri} in a cosmological spacetime
(without event horizon)
\cite{ZelSta,Gri76,HuPar77,HuPar78,FHH,HarHu,Har81,cpcbkr}.
The conceptual framework we adopt is that of a quantum open system \cite{qos},
the formal scheme is that of the closed-time-path \cite{ctp}
and the influence functional \cite{FeyVer,CalLeg83} formalisms, and
the paradigm we use for comparison is that of quantum Brownian motion
\cite{Gra,HPZ1,HPZ2}.

Sciama \cite{Sciama} was the one who had the great insight of proposing
a fluctuation-dissipation relation \cite{CanSci,SCD}
for the depiction of quantum processes in black holes \cite{Haw},
uniformly-accelerated observers \cite{Unr,FulDav}
and de Sitter universe \cite{GibHaw}.
Using Einstein's analysis of the Brownian motion as a guide
he showed that the Hawking and Unruh radiations can be seen as excitations of
vacuum fluctuations and the detector response as following a
dissipation-fluctuation relation. A crucial element for this interpretation
to be possible is the existence of an Euclidean section in the
Schwarzschild, Rindler and de Sitter metrics, imparting a periodicity
in the Green's  function of the matter field,
thus turning it into a thermal propagator \cite{HarHaw76,GibPer}
(in the imaginary-time Matsubara sense),
and forging the equivalence of the system with finite-temperature results
\cite{FulRam}. The same condition applies to Hawking radiance in
de Sitter universe \cite{GibHaw}, which by virtue of its possession of an event
horizon, also admits a FDR interpretation  \cite{Mottola}.
The derivation of the FDR in this class of spacetime was based on a
linear response theory, which hinges on the thermal equilibrium condition
set up by the created particles. For spacetimes without an event horizon,
or for systems under non-equilibrium conditions, one would not ordinarily
think that a FDR could exist \cite{vanKampen,qos}. The generalization of this
relation to non-equilibrium conditions is a much more difficult problem.

When Sciama first proposed this way of thinking, one of us was involved in
the backreaction studies of quantum processes in cosmological spacetimes
\cite{cpcbkr}. The dissipative effect of particle creation on the
dynamics of spacetime seems to point to the existence of a
fluctuation-dissipation relation, except that one factor (dissipation) is
not clear, and the other factor (noise) is missing. Two important
steps had to be taken before this picture began to make better sense.
In the calculation of Hartle and Hu \cite{HarHu} on anisotropy damping,
the Schwinger-DeWitt effective action (`in-out') formalism \cite{SchDeW}
gives  rise to an effective geometry
which is complex, making it difficult to interpret what dissipation really is.
The adoption of the Schwinger-Keldysh (closed-time-path, CTP, or `in-in')
\cite{ctp}
formalism by Calzetta and Hu \cite{CH87} yields a real and causal equation
of motion for the effective geometry, from which one can relate the source of
dissipation in spacetime dynamics to the energy density of particles created
and explicitly identify the viscosity function associated with anisotropy
damping
\cite{CH89}.

The adoption of the CTP formalism was an encouraging step in
the right direction, but one needs to understand the statistical mechanical
meaning of these quantum processes better in order to appraise the validity
of adopting the well-established concepts and results in statistical mechanics
for their depiction or explication. For the particular task of
showing a fluctuation-dissipation relation at work for
quantum fields in a general cosmological spacetime not required to possess
an event horizon, there is also the noise or fluctuation term missing.
These inquiries were summarized in a report written by one of us
\cite{HuPhysica}
in which some tentative replies were given in the form of three conjectures:\\
1) That colored noise associated with quantum field fluctuations is generally
expected in gravitation and cosmology;\\
2) That the backreaction of particle creation in a dynamical spacetime
can be viewed as the manifestation of a generalized fluctuation-dissipation
relation; and \\
3) That all effective field theories, including semiclassical gravity or
even quantum gravity (to the extent that it could be viewed as an effective
field theory), are intrinsically dissipative in nature.\\
There, it was also suggested that one can use the Caldeira-Leggett model
\cite{CalLeg83}
to study the theoretical meaning of dissipation and probe into the relation
of noise and dissipation. The next stage of work in this quest concentrated on
the properties of quantum open systems \cite{qos}
 and extending the theory
to quantum fields and to curved spacetimes.

Using the influence functional formalism of Feynman and Vernon \cite{FeyVer},
one can identify
noise in an environment from the imaginary part of the influence action.
The characteristics of noise depends on the spectral density of the
environment,
the coupling of the system with the environment, and other factors.
Using a model of the Brownian particle coupled
nonlinearly with a bath of harmonic oscillators, Hu, Paz and Zhang
\cite{HPZ1} deduced the noise autocorrelation functions and
a generalized fluctuation-dissipation relation for systems driven by intrinsic
colored and multiplicative noises. This forms the basis for the second stage
of investigation.
To generalize these results to quantum fields, Hu and Matacz \cite{HM2,HMLA}
recently
analyzed the problem of QBM in a parametric oscillator bath.
A parametric oscillator bath enables one to study particle creation in
quantum fields, where the Brownian particle can play the role of
an Unruh detector, or, in a cosmological backreaction problem,
the scale factor of the universe. One can study the detector's response
or the effect on the universe due to the fluctuations of the quantum field.
They found that the characteristics of quantum noise vary
with the nature of the field, the type of coupling between the field and the
background spacetime, and the time-dependence of the scale factor of the
universe.
They showed how a uniformly accelerating
detector in Minkowski space, a static detector outside a black hole
and a comoving observer in a de Sitter universe all observe a thermal spectrum.
By writing the influence functional in terms of the Bogolubov
coefficients which determine the amount of particles produced,
they also identified the origin of noise in this system to particle
creation \cite{nfsg,HM3}.
The influence functional method not only reproduces the known results, but
also enables one to look into the hitherto unknown domain of noise,
fluctuations,
and decoherence.

A program for studying the backreaction of particle creation in
semiclassical cosmology in the open system conceptual framework using influence
functional methods was recently outlined in \cite{HM3,nfsg,Banff}.
The backreaction of these quantum field processes manifests as dissipation
effect, which is described by the dissipation kernel in the influence action.
Using a model where the quantum Brownian particle
and the oscillator bath are coupled parametrically (the field parameters
change in time through the time-dependence of the scale factor of the
universe, which is governed by the semiclassical Einstein equation)
Hu and Matacz \cite{HM3} derived an expression for the influence functional
in terms of the Bogolubov coefficients as a function of the scale factor.
{}From the variation of the influence action they obtained an
equation of motion describing the dynamics of spacetime in the form of an
Einstien-Langevin equation.

After these recent works, it is clear that the influence functional method
is the appropriate framework for studying the nature and origin of noise in
quantum fields and to explore the statistical mechanical meaning of
quantum processes like particle creation and backreaction in the early universe
and black holes. Two additional aspects, however, need be considered to
complete
the story. First, how is it related to the CTP formalism, which gave us,
to begin with, the correct dissipation side of the story? This problem
was taken up in a recent paper of Calzetta and Hu \cite{nfsg},
who showed how noise and fluctuations
in semiclassical gravity can also be obtained with the original CTP formalism.
They also showed that the CTP and the IF formalisms are indeed intimately
related.
They derived an expression for the CTP effective action in terms of the
Bogolubov coefficients and showed how noise is related to the fluctuations
in particle number. From there, they show how an Einstein-Langevin equation
naturally arises as the equation of motion for the effective geometry,
from which a new, extended theory of semiclassical gravity is obtained.
The work of Calzetta, Matacz and the present authors shows clearly that
the old framework of semiclassical gravity  is only a mean field theory.
This theory based on the Einstein equation with a source driven by
the expectation value of the energy momentum tensor should be
replaced by one based on an Einstein-Langevin equation
which describes also the fluctuations of matter fields and dissipative dynamics
of spacetime.

Notice that in moving from the first stage of this investigation based on the
CTP formalism to the second stage based on the IF formalism, one has to elevate
the treatment of classical spacetimes as external fields to
reduced density matrices. In making these transitions and back,
several issues need be addressed. The central issue is the quantum to
classical transition for the spacetime sector \cite{decQC}. The important
question
behind the transition from quantum gravity to semiclassical gravity
is decoherence. This is a subject of much recent interest.
We refer the reader to  recent work for the exposition of different viewpoints
and approaches \cite{envdec,conhis,envhis}. Here,
for the backreaction problem, we shall adopt the results of Paz and Sinha
\cite{PazSin}, which is based on a reduced density matrix formalism adapted to
quantum cosmology. There, the model of a Bianchi-I universe
coupled to a scalar field was used to derive  conditions for
transition from quantum cosmology to the semiclassical limit
via decoherence, and the relationship between decoherence
and backreaction was investigated.

After these previous investigations paved the way for the use of open-system
concepts applied to the backreaction problem of quantum fields
in curved spacetimes, we are finally in a position
to look at the full picture and explore the existence of a
fluctuation-dissipation relation for semiclassical gravity in general.
We shall use the
model of particle creation in Bianchi Type I universe to explore this
relation. In Sec. 2, we give a summary of the results for the quantum Brownian
model, assuming a general nonlinear coupling between the system and the
environment, giving rise to colored and multiplicative noise. Readers familiar
with the QBM problem can skip over this section. In Sec. 3
we begin with the density matrix of the universe and show how
coarse-graining the matter field viewed as an environment produces the
reduced density matrix, and how the influence functional defined  in
the evolutionary operator for the reduced density matrix contains the
relevant information we need-- the dissipation and noise kernels.
In Sec. 4 we analyze the phase and the real components of the influence
functional in detail, sorting out the divergent and renormalized terms in
the phase. We show that the renormalized  phase part provides the dissipative
term in the equation of motion,  and the real component contributes to
decoherence and noise. We show how a colored noise of the quantum field
can be identified, and with it the existence of a fluctuation-dissipation
relation between these kernels. In Sec. 5, we discuss the physical meaning of
this relation. We first show that noise measures the difference in
the amounts of particle creation along two histories. Since this is also the
condition for decoherence to occur, we see that a relation also exists between
decoherence and particle creation. With this noise term, we then
derive the Einstein-Langevin equation for the anisotropy tensor. We show that
it is identical in form to that derived via the CTP formalism before
\cite{CH87}, but
with a new stochastic source term from the noise, as anticipated in
\cite{HuPhysica}. Finally, we show how the dissipation in the anisotropy of
spacetime can be related to the particles created. Thus noise and dissipation
which are connected by a formal relation, are both related to particle
creation,
and the backreaction of particle creation is an embodiment of the FDR.
In Sec. 6 we discuss the physical interpretation of the FDR in a more general
context. We show how the changing rate of  particle creation and
the strength of backreaction effect can be gauged consistently by the
fluctuation-dissipation relation valid for time-dependent conditions.
We also describe related problems for future investigations.

\section{Influence Functional for Quantum Open System}

\subsection{Quantum Brownian Motion Paradigm}

Let us first review a model problem of quantum Brownian motion (QBM)
where the role of noise and dissipation are well understood.
Subsequently we will draw analogies from this problem to analyze
the quantum cosmology problem of our interest.

Consider a Brownian particle interacting with a set of harmonic  oscillators.
The classical action of the Brownian particle is given by
\be
S[x]=\int_0^tds\Bigl\{{1\over 2}M\dot x^2-V(x)\Bigr\}.
\te
The action for the environment is given by
\be
S_e[\{q_n\}]
=\int_0^tds\sum_n\Bigl\{
 {1\over 2}m_n\dot q_n^2
-{1\over 2}m_n\omega^2_nq_n^2 \Bigr\}.
\te
We will assume  that
the action for the system-environment interaction has the following form
\be
S_{int}[x,\{q_n\}]
=\int\limits_0^t ds\sum_n v_n(x) q_n^k  \label{int}
\te

\noindent where  $v_{n}(x) = -\lambda c_nf(x)$ and $\l$ is a
dimensionless coupling constant.
If one is interested only in the averaged effect of the environment on the
system
the appropriate object to study is
the reduced density matrix of the system $\r_r$, which is related to the
full density matrix $\r$ as follows
\be
\rho_r(x,x')
=\int\limits_{-\infty}^{+\infty}dq\int\limits_{-\infty}^{+\infty}
  dq'\rho(x,q;x',q')\delta(q-q').
\te
\noindent It is propagated in time by the evolution operator ${\cal J}_r$
\be
\rho_r(x,x',t)
=\int\limits_{-\infty}^{+\infty}dx_i\int\limits_{-\infty}^{+\infty}dx'_i~
 {\cal J}_r(x,x',t~|~x_i,x'_i,0)~\rho_r(x_i,x'_i,0~).
\te
If we assume that at a given
time $t=0$ the system and the environment are uncorrelated
\be
\hat\rho(0)=\hat\rho_s(0)\times\hat\rho_e(0),
\te
then ${\cal J}_r$ does not depend on the initial state of the system
and can be written as
\bea
{\cal J}_r(x_f,x'_f,t~|~x_i,x'_i,)
& =& \int\limits_{x_i}^{x_f}Dx                     \label{prop}
   \int\limits_{x'_i}^{x'_f}Dx'~
   \exp{i\over \hbar}\Bigl\{S[x]-S[x']\Bigr\}~{\cal F}[x,x']  \nn \\
& =& \int\limits_{x_i}^{x_f}Dx
   \int\limits_{x'_i}^{x'_f}Dx'~
   \exp{i\over \hbar} S_{eff}[x,x']
\tea
where the subscripts  $i,f$ denote initial and  final variables,
and $S_{eff}[x,x']$ is the effective action for the open quantum system.
The influence functional ${\cal F}[x,x']$ is defined as
\bea
 {\cal F}[x,x'] & = &\int\limits_{-\infty}^{+\infty}dq_f
 \int\limits_{-\infty}^{+\infty}dq_i
 \int\limits_{-\infty}^{+\infty}dq'_i          \label{ifbm}
 \int\limits_{q_i}^{q_f}Dq
  \int\limits_{q'_i}^{q_f}Dq' \nn \\
& & \times \exp{i\over\hbar}\Bigl\{
  S_e[q]+S_{int}[x,q]-S_e[q']-S_{int}[x',q'] \Bigr\}
  \rho_e(q_i,q'_i,0) \nn \\
& = & \exp{i\over\hbar} S_{IF}[x,x']
\tea
where $S_{IF}[x,x']$ is the influence action.  Thus $ S_{eff}[x,x'] =
S[x]-S[x'] + S_{IF}[x,x']$.

{}From its definition it is obvious that if the interaction term is zero, the
influence functional is equal to unity and the influence action is zero.
In general, the influence functional
is a highly non--local object. Not only does it depend on the time history,
but --and this is the more important property-- it also
irreducibly mixes the two sets
of histories in the path integral of (2.7).
Note that the histories
$ x $ and $ x' $ could be interpreted as moving
forward and backward in time respectively.
Viewed in this way, one can see the similarity of the influence functional
and the generating functional in the closed-time-path, or Schwinger-Keldysh
\cite{ctp} integral formalism.

We will assume that initially the bath is in thermal equilibrium at a
temperature $T = {(k_B \beta)}^{-1}$. The $T = 0 $ case corresponds to the
bath oscillators being in their respective ground states.
It can be shown \cite{HPZ2} that the influence action for the model given
by the interaction in (2.3)  to second order in $\lambda$ is given
by
\bea
& &S_{IF} [x,x']
= \Bigl\{\int\limits_0^tds~[-\Delta V(x)~]
  -\int\limits_0^tds~[-\Delta V(x')]~\Bigr\} \nn \\
& & -\int\limits_0^tds_1\int\limits_0^{s_1}ds_2~\lambda^2 \label{iabm}
   \Bigl[f(x(s_1))-f(x'(s_1))\Bigr]\mu^{(k)}(s_1-s_2)
   \Bigl[f(x(s_2))+f(x'(s_2))\Bigr]  \nn \\
& &+i\int\limits_0^tds_1\int\limits_0^{s_1}ds_2~\lambda^2
   \Bigl[f(x(s_1))-f(x'(s_1))\Bigr]\nu^{(k)} (s_1-s_2)
   \Bigl[f(x(s_2))-f(x'(s_2))\Bigr]
\tea
where $\Delta V(x)$ is a renormalization of the potential that arises
from the contribution of the bath. It appears only for even $k$ couplings.
For the case $k=1$ the above result is exact.
This is a generalization of the
result obtained in \cite{FeyVer} where it was shown that the
non-local kernel $\mu^{(k)}(s_1-s_2)$ is associated with dissipation or
the generalized viscosity function that appears in the corresponding
Langevin equation and $\nu^{(k)}(s_1-s_2)$ is associated with the
time correlation function of the stochastic noise term. The dissipation
part has been studied in detail  by Calzetta, Hu, Paz, Sinha and others
\cite{CH87,CH89,disQC,SinHu,PazSin,nfsg}
in cosmological backreaction problems. We shall elaborate somewhat on
the nature of the noise part of the problem and then analyze their connection.
In general $\nu$ is nonlocal, which gives rise to colored noises.
Only at high temperatures would the noise kernel become a delta function,
which corresponds to a white noise source. Let us first see the meaning
of the noise kernel.

\subsection{Noise}

The noise part of the influence functional is given by
\be
\exp\{ -{1\over \hbar}\int\limits_0^tds_1\int\limits_0^{s_1}ds_2~
   \Bigl[f(x(s_1))-f(x'(s_1))\Bigr]\nu^{(k)}(s_1-s_2) \label{expnoise}
   \Bigl[f(x(s_2))-f(x'(s_2))\Bigr]
\te
where $\nu^{(k)}$ is redefined by absorbing the $\l^2$.
This term can be rewritten using the following functional Gaussian identity
\cite{FeyVer}
which states that the above expression is equal to
\be
\int {\cal D}\xi^{(k)}(t) {\cal P}[\xi^{(k)}]\exp\{ {i\over
\hbar}\int\limits_0^t
ds \xi^{(k)}(s)[f(x(s)) - f(x'(s))]
\te
where
\be
{\cal P}[\xi^{(k)}] = P^{(k)} \exp\{-{1\over        \label{noisedis}
\hbar}\int\limits_0^tds_1\int\limits_0^tds_2 \ha \xi^{(k)}(s_1)
[\nu^{(k)} (s_1 - s_2)]^{-1}\xi^{(k)}(s_2) \}
\te
is the functional distribution of $\xi^{(k)}(s)$ and $P^{(k)}$ is a
normalization factor given by
\be
[P^{(k)}]^{-1} = \int{\cal D}\xi^{(k)}(s)\exp\{-{1\over \hbar}
\int\limits_0^tds_1\int\limits_0^tds_2\xi^{(k)}(s_1)[\nu^{(k)}(s_1-s_2)]^{-1}
\xi^{(k)}(s_2)\}.
\te
The influence functional can then be rewritten as
\bea
{\cal F}[x,x'] &=& \int{\cal D}\xi^{(k)}(s){\cal P}[\xi^{(k)}] exp{i\over
\hbar}
\hat S_{IF}[x,x',\xi^{(k)}]\nn \\
&\equiv& {\left\langle \exp {i\over \hbar}\hat
S_{IF}[x,x',\xi^{(k)}]\right\rangle}_{\xi}
\tea
where
\bea
& &\hat S_{IF}[x,x',\xi^{(k)}]  =   \int\limits_0^tds~\Bigl\{-\Delta
V(x)~\Bigr\}
  -\int\limits_0^tds~\Bigl\{-\Delta V(x')~\Bigr\} \nn \\
& & -\int\limits_0^tds_1\int\limits_0^{s_1}ds_2~
   \Bigl[f(x(s_1))-f(x'(s_1))\Bigr]\mu^{(k)}(s_1-s_2)  \label{ianoise}
   \Bigl[f(x(s_2))+f(x'(s_2))\Bigr]  \nn \\
& & - \int\limits_0^t ds \xi^{(k)}(s)f(x(s))
+\int\limits_0^tds\xi^{(k)}(s)f(x'(s))
\tea
so that the reduced density matrix can be rewritten as
\be
\rho_r(x,x') = \int{\cal D}\xi^{(k)}(s){\cal
P}[\xi^{(k)}]\rho_r(x,x',[\xi^{(k)}]).
\te

The full effective action can be written as
\bea
S_{eff}[ x, x', \xi ] &=& \{ S[x] + \int\limits_0^tds~\xi(s) f(x(s))\} - \{
S[x'] + \int\limits_0^tds~\xi(s) f(x'(s))\} \nn \\
                      &- & \int\limits_0^tds_1\int\limits_0^{s_1}ds_2~
   \Bigl[f(x(s_1))-f(x'(s_1))\Bigr]\mu^{(k)}(s_1-s_2)
   \Bigl[f(x(s_2))+f(x'(s_2))\Bigr]
\tea

{}From equation (2.15) we can view $\xi^{(k)}(s)$ as a
nonlinear external stochastic force and the
reduced density matrix is calculated by taking a stochastic average
over the distribution $P[\xi^{(k)}]$ of this source.

{}From (2.12), we can see that the distribution functional is Gaussian.
The Gaussian noise is therefore completely characterized by
\bea
{\langle\xi^{(k)}(s)\rangle}_{\xi^{(k)}} &=& 0 \nn\\
{\langle\xi^{(k)}(s_1)\xi^{(k)}(s_2)\rangle} &=& \hbar\nu^{(k)}(s_1-s_2).
\label{noisecor}
\tea
We see that the non-local kernel $\nu^{(k)}(s_1-s_2)$ is just the two-point
time correlation function of the external stochastic source
$\xi^{(k)}(s)$ multiplied by $\hbar$.

In this framework, the expectation value of any quantum mechanical
variable $Q(x)$ is given by \cite{Zhang}
\bea
\langle Q(x)\rangle & = & \int{\cal D}\xi^{(k)}(s){\cal
P}[\xi^{(k)}]\int\limits_{-\infty}
^{+\infty}dx \rho_r(x,x,[\xi^{(k)}])Q(x) \nn \\
& = & {\left\langle {\langle Q(x)\rangle}_{quantum}\right\rangle}_{noise}.
\tea
This summarizes the interpretation of $\nu^{(k)}(s_1-s_2)$ as a noise or
fluctuation kernel.

\subsection{Langevin Equation}

We now derive the semiclassical equation of motion generated by
the influence action (2.9). This will allow us to
see why the
kernel $\mu^{(k)}(s_1-s_2)$ should be associated with dissipation.
Define a ``center-of-mass" coordinate $\bar x$ and a
``relative" coordinate $\Delta$ as follows
\bea
\bar x(s) & = & {1\over 2}[x(s) + x'(s)] \nn\\
\Delta(s) & = & x'(s) - x(s).
\tea
The semiclassical equation of motion for $\bar x$ is derived by demanding
(cf. \cite{CH87})
\be
\frac{\delta}{\delta\Delta}\Bigl[S[x]-S[x']+S_{IF}[x,x']\Bigr]
\bigg|_{\Delta=0}=0.
\te
Using the sum and difference coordinates (2.20) and the influence action
(2.9) we find that (2.21) leads to
\be
\frac{\partial L_r}{\partial \bar x}-\frac{d}{dt}
\frac{\partial L_r}{\partial \dot{\bar x}} - 2\frac{\partial f(\bar
x)}{\partial \bar x}
\int\limits_0^t ds~\gamma^{(k)}(t-s)\frac{\partial f(\bar x(s))}{\partial s}
= F_{\xi^{(k)}}(t)
\te
where $\frac{d}{ds}\gamma^{(k)}(t-s)=\mu^{(k)}(t-s)$.
We see that this is in the form of
a classical Langevin equation with a nonlinear stochastic force
$F_{\xi^{(k)}}(s) = -\xi^{(k)}(s) \frac{\partial f(\bar x)}{\partial \bar x}$.
This corresponds to a multiplicative noise unless $f(\bar x)=\bar x$
in which case it is additive. $L_r$ denotes a renormalized
system Lagrangian. This is obtained by absorbing a surface term
and the potential renormalization in the influence action into the system
action.
The nonlocal kernel $\gamma^{(k)}(t-s)$ is responsible for non-local
dissipation.
In special cases like a high temperature ohmic environment,
this kernel becomes a delta function and hence the dissipation is local.

\subsection{Fluctuation-Dissipation Relation}

Recall that the label $k$ is the order of the bath variable to which the system
variable is coupled.
$\gamma^{(k)}(s)$ can be written as a sum of various contributions
\be
\gamma^{(k)}(s)=\sum_l\gamma^{(k)}_l(s)
\te
where the sum is over even (odd) values of $l$ when $k$ is even (odd).
To derive the explicit forms of each dissipation kernel, it is useful to
define first the spectral density functions
\be
I^{(k)}(\omega)
= \sum_n~\delta(\omega-\omega_n)~k~\pi\hbar^{k-2}~
   {\lambda^2 c_n^2(\omega_n)\over (2m_n\omega_n)^k }.
\te
It contains the information about the environmental mode density and coupling
strength as a function of frequency.
Different environments are classified according to the
functional form of the spectral density $I(\o)$.


In terms of these functions, the dissipation kernels can be written
as
\be
\gamma^{(k)}_l(s)
 =\int\limits_0^{+\infty}{d\omega\over\pi}
 ~{1\over \omega}I^{(k)} (\omega)
 ~ M^{(k)}_l(z)~\cos l\omega s
\te
where $M^{(k)}_l(z)$ are temperature dependent factors derived in \cite{HPZ2}
and $ z = coth {\ha \beta \hbar \omega}$.
Analogously, the noise kernels $\nu^{(k)}(s)$
can also be written as a sum of various contributions
\be
\nu^{(k)}(s) = \sum_l \nu^{(k)}_l(s)
\te
\noindent where the sum runs again over even (odd) values of $l$ for
$k$ even (odd). The kernels $\nu^{(k)}_l(s)$ can be written as
\be
\nu^{(k)}_l
=\hbar\int\limits_0^{+\infty}{d\omega\over\pi}~
 I^{(k)} (\omega)~N^{(k)}_l(z)~\cos l\omega s
\te
where $ N^{(k)}_l (z)$ is another set of temperature- dependent factors given
by
\cite{HPZ2}

To understand the physical meaning of the noise kernels of different orders,
we can think of them as being associated with  $l$ independent stochastic
sources that are coupled to the Brownian particle through interaction
terms of the form (2.15)
\be
\int\limits_0^tds~\sum_l~\xi_l^{(k)}(s)~f(x).
\te
\noindent This type of coupling generates a stochastic force in the associated
Langevin equation
\be
F_{\xi_l^{(k)}}(s)=-\xi_l^{(k)}(s)\frac{\partial f(x)}{\partial x}
\te
\noindent which corresponds to multiplicative noise.
The stochastic sources $\xi_l^{(k)}$ have a probability distribution given by
(\ref{noisedis})
which generates the correlation functions (\ref{noisecor}) for each $k$ and
$l$.

To every stochastic source we can associate a dissipative term that is
present in the real part of the influence action. The dissipative and the
noise
kernels are related by generalized fluctuation--dissipation relations
of the following form
\be
\nu^{(k)}_l(t)
=\int\limits_{-\infty}^{+\infty}ds~K^{(k)}_l(t-s)~\gamma^{(k)}_l(s)
\label{FDR}                                                        
\te
\noindent where the kernel $ K^{(k)}_l(s) $ is
\be
K^{(k)}_l(s)
=\int\limits_0^{+\infty}{d\omega\over\pi}~
  L^{(k)}_l(z)~l~\omega~\cos~l\omega s
  \label{K}                                                        
\te
and the temperature-dependent factor $L^{(k)}_l(z)= N^{(k)}_l(z)/
M^{(k)}_l(z)$.

A fluctuation dissipation relation of the form (\ref{FDR}) exists for
the linear case where the temperature dependent factor appearing in (\ref{K})
is simply $L^{(1)}=z$. The fluctuation-dissipation kernels $K_l^{(k)}$
have rather complicated forms except in some special cases.
In the high temperature limit, which is characterized
by the condition
$ k_BT\gg \hbar\Lambda $, where $ \Lambda $ is the cutoff frequency of
the environment, $z=\coth \beta\hbar\omega/2
\to 2/\beta\hbar\omega$                                      
we obtain
\be
L^{(k)}_l(z) \to {{2k_BT}\over {\hbar\omega}}.
\te
\noindent In the limit $ \Lambda \to +\infty $, we get the general result
\be
K^{(k)}_l(s)= {2k_BT\over \hbar}\delta(s)
\te
\noindent which tells us that at high temperature there is
only one form of fluctuation-dissipation relation, the Green-Kubo relation
\cite{fdr}
\be
 \nu^{(k)}_l(s)
={2k_BT\over \hbar}\gamma^{(k)}_l(s).
\te
\noindent In the zero temperature limit, characterized by $~ z \to 1,~ $
we have
\be
L^{(k)}_l(z) \to l.
\te
\noindent The fluctuation-dissipation kernel becomes $k$-independent
and hence identical to the one for the linearly- coupled case
\be
K(s)
=\int\limits_0^{+\infty}
 {d\omega\over\pi}~\omega\cos\omega s. \label{zerofd}
\te

It is interesting to note that the fluctuation-dissipation relations for the
linear and the nonlinear dissipation models are exactly identical both
in the high temperature and in the zero temperature limits. In other words,
they are not very sensitive to the different
system-bath couplings at both high and zero temperature limits.
The fluctuation-dissipation relation reflects a
categorical relation (backreaction) between the stochastic stimulation
(fluctuation-noise) of the environment and the averaged response of a system
(dissipation) which has a much deeper and universal meaning than that
manifested
in specific cases or under special conditions.

Our aim in the next section would be to consider a model consisting
of quantum fields coupled to a cosmological background metric and
cast it into the system-environment form as discussed here.
Consequently we shall see that one can construct an influence
functional of a form very similar to (\ref{ifbm}) and hence derive a
fluctuation-dissipation relation of the form (\ref{FDR}).

\section{Influence Functional for Quantum Cosmology}

\subsection{Reduced Density Matrix of the Universe}

The model we will analyze here is the same as that used in
\cite{PazSin} from which we will quote results relevant to our study.
Our ``system"  will consist of a minisuperspace model
with $D$ degrees of freedom denoted by coordinates
$r^m$ (with $m=1,\ldots,D$). The minisuperspace modes will be coupled to
``environment" degrees of freedom that we schematically represent by $\Phi$
(they
will be later
associated with the modes of a scalar field). The quantum
mechanical description of this Universe will be given by the wave function
of the Universe  $\Psi=\Psi(r^m, \Phi)$ which,
as a consequence of the existence of a classical Hamiltonian constraint,
satisfies the Wheeler- DeWitt equation:
\be
H\Psi= \bigl(H_r + H_\Phi\bigr)\Psi=0   \label{WD}
\te
\noindent In the class of  models we consider, the Hamiltonian corresponding to
the minisuperspace variables can
be written as
\be
H_r= {1\over{2M}}G^{mm'}p_mp_{m'} + M V(r^m)  \label{hamiltonian}
\te
The matrix $G^{mm'}$ determines the metric in the minisuperspace
(the supermetric) and the quantity $M$ is proportional to the
square of the Planck mass. In the following we will set $\hbar = 1$ throughout.
In the above Wheeler- DeWitt equation we assume that the momenta are
replaced by operators according to a covariant factor ordering
prescription.
The Hamiltonian constraint represents an important distinction from
the quantum Brownian motion case discussed previously, because it
implies that there is no preferred notion of time in this case and
the wavefunction satsfies (\ref{WD}) rather than the Schr\"odinger
equation.
The Hamiltonian associated with the environment degrees of
freedom is some function $H_\Phi(\Phi, \pi_\Phi, r^m, p_m)$ that we
will specify later.

We will be interested in making predictions concerning only the
behavior of the minisuperspace variables $r^m$ which we consider
the ``relevant'' part of the universe. To achieve such a
coarse- grained description we will work with the reduced density matrix
of the system which is defined as:
\be
\rho_{red}(r',r) = \int d\Phi \Psi^{*}(r,\Phi) \Psi(r',\Phi)
\te
For some region of the minisuperspace,
(\ref{WD}) admits solutions that are oscillatory functions of $r^m$ of the
following WKB form:
\be
\Psi(r,\Phi) = e^{iMS(r)}C(r)\psi(r,\Phi)  \label{WKB}
\te
In this regime, the system variables $r$ and the environment variables
$\Phi$ behave as heavy and light modes respectively (the Planck mass
plays the role of a large mass parameter) in analogy  with the
Born-Oppenheimer approximation. This also provides some
justification of the system-environment split akin to the Brownian
motion case . Thus
if one assumes that all the functions $S,C,\psi$ can be
expanded in powers of $M^{-1}$ and substitutes these expansions into
(\ref{WD}), one gets, to leading order (i.e., $M^0$):
\be
{1\over 2} G^{mm'}{{\partial S_0}\over{\partial r^m}} \label{HJ}
{{\partial S_0}\over{\partial r^{m'}}} + V(r) = 0
\te
\noindent which is essentially the minisuperspace version of
the Hamilton-Jacobi equation.
To the next order in $M$ one obtains,
\be
iG^{mm'}{{\partial S_0}\over{\partial r^m}} {{\partial}\over{\partial r^{m'}}}
\psi_0= H_\Phi(\Phi, \pi_\Phi, r^m, p_m={{\partial S_0}\over
{\partial r^m}})\psi_0  \label{Schr}
\te
\noindent This last equation is obtained provided we choose the prefactor
$C_0$ identical to the $H_\Phi=0$ case.
Thus, if we define the WKB time $t$ as
\be
{d\over dt} = G^{mm'}{{\partial S_0}\over{\partial r^{m'}} }
\label{time}
{{\partial}\over{\partial r^m}}
\te
\noindent the equation (\ref{Schr}) reduces to the familiar Schr\"odinger
equation
that reads:
\be
i{{d\psi}\over {dt}} = H_\Phi \psi
\te
{}From now on we will drop all the $0$-subindices which should be considered
as implicit in all the equations where $S, C$ and $\psi$ appear.
The Hamilton-Jacobi equation (\ref{HJ})
will have a $D-1$ parameter family of solutions and for
each one of these solutions
we can build a wave function like (\ref{WKB}). In general one can  assume that
the wave function
of the Universe is a superposition of these terms, each  of which
will be called a WKB branch:
\be
\Psi(r,\Phi) = \sum\limits_n e^{iMS_{(n)}(r)}C_{(n)}(r)\psi_{(n)}(r,\Phi)
\label{super}
\te
Here the subindex $(n)$ labels the WKB branch characterized by a set of
parameters $(n)$ that uniquely defines the
particular solution to the Hamilton-Jacobi equation. However, in the rest of
our
analysis , we will consider the wavefunction to be represented by a single term
of the above sum, i.e, by a particular WKB branch. We will drop the subscript
$n$ from now on with this understanding.

The reduced density matrix associated with the wave function (\ref{WKB}) is:
\be
\rho_{red}(r',r) = e^{iM[S(r) - S(r')]}
    C(r)C(r') {\cal I}(r',r)
\te

\noindent where
\be
{\cal I}(r',r)= \int \psi^{*}(r',\Phi)\psi(r,\Phi) d\Phi
\te
The influence of the environment on the system is summarized by the
above function ${\cal I}$ and it will be the basic object of our
interest. It has been shown in references \cite{PazSin,Kiefer} that this is
the object that is exactly analogous to the Feynman-Vernon influence
functional ${\cal F} (x, x')$ in the case where the environment
is in a pure state. We
will therefore call ${\cal I}(r',r)$ the influence functional and
analyze the fluctuation and dissipation phenomena in analogy to
the QBM problem.
To facilitate making these connections, we write the
influence functional in the form
\be
{\cal I}(r,r')= \exp\{i\Gamma(r,r')\}
\te
where the influence action can be written as
\be
\Gamma(r, r')
= \Theta(r,r') + i\tilde\Gamma(r,r') \label{ifqc}
\te
The  phase $\Theta$ and the real exponent $\tilde\Gamma$
which constitute the influence functional will be the basic objects of our
interest
(note
that $\tilde\Gamma$ is positive since the overlap is bounded by unity).

\subsection{Bianchi-I Minisuperspace with a Conformal Scalar Field}
We now  specialize our model to a  minisuperspace of Bianchi I universe coupled
to a massless
conformal scalar field. The line element is given by \cite{mss}
\be
ds^2 = a^2   d{\eta}^2 - a^2 e^{2\beta}_{ij} dx^i dx^j ,
\te
where  $\eta$ is the conformal time .
The traceless $3\times 3$ matrix $\beta$ measures the anisotropy, its time rate
of change gives the shear.  For Type-I universe, it can always be
parametrized by the principal eigenvalues
\be
\beta = {\rm diag} (\beta_1, \beta_2, \beta_3)
\te
or, equivalently by  $\beta_\pm$ defined by
\be
\beta_1= \beta_++\sqrt 3 \beta_-,~~ \beta_2=\beta_+ - \sqrt 3 \beta_-,~~
              \beta_3= - 2\beta_+     \label{bpm}
\te

Rewriting the scale factor as $a=e^{\alpha}$,
the Einstein Hilbert action can be written as
\be
S_g = { 6M} \int d\eta
    \{ {e^{2\alpha}} ( -\dot\alpha^2 + \dot\beta_+^2 + \dot\beta_-^2)
\te
where  $M=M_{Pl}^2$ , and a dot denotes taking a
derivative with respect to the conformal time $\eta $.
We normalize the spatial volume to $1$  assuming $T^3$ spatial
topology.

The action for the scalar field is given by
\be
S_f= {1\over 2} \int d^4x ~(g^{\mu\nu}\partial_\mu\Phi\partial_\nu\Phi
                       - {1\over 6} R\Phi^2)
\te
\noindent which, after integrating by parts and defining the conformal field
$X=a\Phi$, can be written as:
\be
S_f= {1\over 2} \int d^4x~ \{ {\dot X^2} +  X \nabla^{(3)}X -
                         (\dot\beta_+^2 + \dot\beta_-^2)X^2\}
\te
\noindent where the spatial Laplacian is given by
$\nabla^{(3)}=e^{2\beta}_{ij}\partial_i\partial_j$.

As usual, we
expand the field $X$ in an orthonormal basis of eigenfunctions of
$\nabla^{(3)}$. As the spatial sections are flat, the
eigenfunctions are simple trigonometric functions and the momenta are
quantized due to the periodic boundary conditions associated with the $T^3$.
We will denote the basis as
$\{Q_{\vk\sigma}(\vec x), {\vk}=(k_x, k_y, k_z),
k_j=2\pi n_j, \sigma=\pm\}$. The index $\sigma$ labels the functions according
to their parity. The expansion of the field $X$ reads:
\be
X(\vec x, \eta) = \sum_{\vk\sigma} Q_{\vk\sigma}(\vec x) \chi_{\vk\sigma}(\eta)
\te

The variables of the minisuperspace constituting our open system
are $r^m=(\alpha,\beta_+, \beta_-)$ or $(\alpha, \beta_{ij})$
and the `environment' variables
are the collection of field amplitudes $\{\chi_{\vk\sigma}, ~~
k_j=2\pi n_j, ~~\sigma=\pm\}$.
Using our previous expressions it is easy to show that the Hamiltonian can
be written in the form of (\ref{hamiltonian}), where the gravitational part
has the supermetric
\be
G^{mm'} = {1\over{a^2}} {\rm diag}(-1,+1,+1).
\te
\noindent On the other hand the matter Hamiltonian can be written
as
\bea
 H_X&=& \sum_{\vk\sigma} H_{\vk\sigma}= \sum_{\vk\sigma}
{1\over 2} (\pi_{\vk\sigma}^2 + \Omega_{\vk}^2 {\chi_{\vk\sigma}}^2 )\nn \\
\Omega_{\vk}^2 &=& e^{2\beta}_{ij} k^ik^j~ +
{1\over {144 M^2~a^4}} (p_{\beta_+}^2 + p_{\beta_-}^2)
\tea

We will assume the wave function of the universe can be
written as (\ref{WKB}), where the function $S$ obeys the
Hamilton--Jacobi equation (\ref{HJ}) which in this case is given by:
\be
{{e^{-2\alpha}}\over{2M}} (-(\partial_\alpha S)^2 + (\partial_{\beta_+} S)^2 +
(\partial_{\beta_-} S)^2 ) = 0  \label{HJbi}
\te
\noindent
This equation can be  separated and solved as
\be
S(\a,\beta_{\pm}) = \tilde S_{\vec b} (\a) + b_+\beta_+ + b_-\beta_-
\label{S}
\te
with
\be
{\partial_\a}\tilde S_{\vec b}(\a) = \pm |\vec b|
\te
\noindent where we use $\vec b$ to denote the two dimensional constant
vector $(b_+, b_-)$.

As we can see,
a particular solution to the Hamilton--Jacobi equation is parametrized
by two integration constants ($b_+$ and $b_-$) and by the sign that
defines $\tilde S(a)$ in
equation (\ref{S}). Therefore, the label $(n)$ that characterizes a
solution of (\ref{HJbi})  stands
for the set of constants $\{\vec b, \pm\}$. Every function
$S_{(n)}$ generates a $2$--parameter family of trajectories in the three
dimensional minisuperspace
(these are the curves orthogonal to the $S_{(n)}=$constant hypersurfaces).
These trajectories are exact solutions to the Einstein's
equations,
and if we restrict our  considerations to the plane
 $(a,\beta_+)$,
the trajectories are defined by the equation
\be
{\partial\beta_+\over{\partial\alpha}} = - {b_+\over {\partial_{\alpha} S}}
\te

\noindent The minisuperspace trajectories can be found by integrating the
above equation and are  straight lines (with slope given by $\pm b_+/|\vec b|$)
corresponding to the well known Kasner's solutions. In that case, for the
``expanding'' (i.e. $\dot\a>0$) branch, we have
$\bpl={{b_+}\over{|\vec b|}}\a+{\bpl}_0$, where ${\bpl}_0$
is an integration constant. 

\subsection{Influence Action}

 We have to compute
the influence functional (\ref{ifqc}) according to
the strategy described in the beginning of this section
and for that  we have to solve
the Schr\"odinger equation (\ref{Schr}).
It is possible to make the following ansatz for the matter wave function
\be
\psi(r,X)=
\psi(r, \{\chi_{\vk}\}) = \prod_{\vk} {\psi}_{\vk}(r, \chi_{\vk})
\te
Thus, the influence functional is expressed as an infinite product while
the phase $\Theta$ and the
real exponent $\tilde\Gamma$ can be written as a sum  of contributions from
each mode.

Each component of the wave function
satisfies the following Schr\"odinger equation:
\be
i{{\partial \psi_{\vk}}\over{\partial \eta}} = {H}_{\vk}~ \psi_{\vk}.
\te
\noindent with a Hamiltonian given by :
\bea
H_{\vk} &=& -{1\over 2}{{d^2~}\over{d\chi_{\vk}^2}} + {1\over 2}\Omega_{\vk}^2
\chi_{\vk}^2\nn \\
&=& -{1\over 2}{{d^2~}\over{d\chi_{\vk}^2}} +
{1\over 2} (e^{2\beta_{ij}}~k^i k^j +
{\dot\beta_+}^2+{\dot\beta_-}^2){\chi_{\vk}^2}
\tea
\noindent where as before, we have used a dot to denote the derivative with
respect to the
conformal time, which also happens to coincide wth the WKB time as can be seen
from
applying the definition (3.7) to the model of sect. 3.2.

Let us now describe how  we compute the influence functional.
We will make a Gaussian
ansatz for the wave function $\psi_{\vk}$ that corresponds
to assuming that the state for the scalar perturbations is a particular
vacuum .
Thus, we write each component of the wave function as (for
simplicity we will omit the index ${\vk}$):
\be
\psi (r, f) = ({\pi\over w_i})^{1\over 4} ~e^{-{i\over 2}\int w_idt}
{}~e^{{i\over 2}f^2 w}
\te
where $w \equiv \dot u/ u \equiv w_r + i w_i$, and $w_r, w_i$ are the
real and imaginary parts of $w$.
The  equation satisfied by the function $u$ is easily
derived from the Schr\"odinger equation and can be written  as :
\be
\ddot u + \Omega_{\vk}^2 u = 0        \label{weq}
\te

The computation of the overlap factor
involves solving the above equations.
In our model this can be done using a
perturbative scheme if we assume that the anisotropy coordinates
are small. In that case, we can can write (up to second order in the
anisotropy):
\be
\Omega_{\vk}^2 = \ok^2 - \lambda_1 - \lambda_2 \label{omega}
\te
where
\bea
\omega_k & = &|{\bf k}^2|^{1/2}, \quad
\lambda_1 = ~ - 2 \beta_{ij} k^ik^j \quad{\rm and}\quad \nn \\
\label{lambda}
\lambda_2 & =& -~2\beta^2_{ij}k^ik^j - (\dot\bpl^2 + \dot\bmn^2)
\tea
Then, the equation for $u$ can be solved by a standard iteration procedure
\cite{PazSin,Unr}.
Assuming that the anisotropy is ``switched off" at early and late
times, and taking  the
initial state as the  conformal
vacuum, the expressions for $\tilde\Gamma$ and $\Theta$
of the exponent of the influence functional defined in (\ref{ifqc})
are given respectively by
\bea
\tilde\Gamma(r,r') & = &
{\ok^2} cos(2\ok(\eta_1-\eta_2))  \nn \\
& & +{1\over 16} \int^{\eta'}\int^{\eta_1} d\eta_1~d\eta_2
{{\lambda_1(\eta_1)\lambda_1(\eta_2)} \over {\ok^2}}
cos(2\ok(\eta_1-\eta_2)) \nn \\    \label{rexp}
& &-{1\over 16} \int^\eta\int^{\eta'} d\eta_1~d\eta_2
{{\lambda_1(\eta_1)\lambda'_1(\eta_2)}\over {\ok^2}}
cos(2\ok(\eta'-\eta+\eta_1-\eta_2))
\tea
and
\bea
\Theta(r,r') & = & ~ {1\over 2}\ok(\eta-\eta') + {1\over{4\ok}}
\int^{\eta}d\eta_1 \lambda_2(\eta_1) - {1\over{4\ok}} \int^{\eta'}d\eta_1
\lambda_2(\eta_1)~+\nn \\   \label{phase}
 & + & {1\over 8} \int^\eta\int^{\eta_1} d\eta_1~d\eta_2 ~
{{\lambda_1(\eta_1)\lambda_1(\eta_2)}\over
{\ok^2}} ~sin(2\ok(\eta_1-\eta_2)) ~- \nn \\
& - & {1\over 8} \int^{\eta'}\int^{\eta_1} d\eta_1~d\eta_2 ~{{\lambda_1(\eta_1)
\lambda_1(\eta_2)} \over  {\ok^2}} ~sin(2\ok(\eta_1-\eta_2)) ~+\nn \\
& + &{1\over 8} \int^\eta\int^{\eta'} d\eta_1~d\eta_2 ~{{\lambda_1(\eta_1)
\lambda_1(\eta_2)}\over {\ok^2}} ~sin(2\ok(\eta'-\eta+\eta_1-\eta_2))
\tea
up to second order in anisotropy.
The total phase $\Theta$ and the total real exponent $\tilde \Gamma$
of the influence functional are obtained by summing over $\vk$ of
(\ref{phase}) and (\ref{rexp}) respectively.
In performing these sums, divergent expressions will arise which
will have to be regularized and  renormalized.

The above equations clearly show the history
dependence of the influence functional since they are written in terms of time
integrals of functions that depend on $\bpm(\eta_1)$ . Therefore, the phase and
the real exponent are {\it functionals} of the zero order WKB histories.

Notice that since in this model  we have more than
one minisuperspace degree of freedom, even within a WKB branch,
we have a whole family of trajectories rather than a
single trajectory. So as far as the solutions of the Hamilton-Jacobi
equation is concerned, this implies restricting ourselves to the
family of trajectories  given by the solution of  ( 3.26)
with a fixed value of ${b_+/ |{\vec b}|}$. These are
 a family of parallel
straight lines with the slope fixed by $n$ and different $\beta$
intercepts. We note that in the configuration space of the $\alpha-\beta_+$
plane, one and only one trajectory passes through each
point. Hence each point in configuration space can be associated with
an entire history, and thus ${\cal I}$ is a functional of two histories
as in the Brownian motion example.

As it stands, ${\cal I}(r,r')$ is still not in a form that can be
put in one-to-one correspondence with the ${\cal F}(x,x')$ of the QBM problem,
because the latter is an explicit function of time, whereas in the
former, the WKB time is defined through (\ref{time}) as a function of the
coordinates $r$. The definition of $\eta$= constant surfaces
depends on the choice of the hypersurface in minisuperspace on which
the initial condition of the wave function is specified. In our case the
initial
conditions were specified on a  $\alpha$ = constant hypersurface. Thus
our constant WKB time hypersurfaces are those with $\alpha$=
constant. Now, let us specialize to the situation where ${\cal I}(r,r')$
is evaluated on two points such that $\a = \a'$. From the above
discussion then we know that this implies that $\eta = \eta'$.
The two histories, $\bpm(\eta_1)$ and
$\beta'_\pm(\eta_1)$ that enter into the calculation of the influence
functional
are the parallel lines (with slope determined by $(n)$) , passing through
the points $(\alpha, {\beta}_+)$ and $(\alpha, {{\beta}_+}')$ respectively.
Now, the influence functional can be written as ${\cal I}(\bpm,{\bpm}',\eta)$
and can finally be  compared  with that of the QBM problem.


\section{Fluctuations in Quantum Fields and Dissipation of Spacetime
Anisotropy}

\subsection{Regularized Influence Action}

 It has been pointed out in \cite{PazSin} that the
influence action $\Gamma$ is identical to the
Schwinger--Keldysh (or Closed Time Path) effective action which is a
functional of two histories and can be computed using diagrammatic
techniques. Thus $\Gamma$ is esentially the same as the quantity
given by  (3.11 ) in \cite{CH87}, with $\beta$ and $\beta'$ corresponding
to ${{\beta}_{ij}}^+ , {{\beta}_{ij}}^{-}$ in the CTP context, where the $+$
and $-$ superscripts refer to the positive and negative contour branches
respectively.
This identification is useful as it connects with the well-known
results in semiclassical gravity \cite{CH87}.
This connection provides both conceptual and technical advantages as it
offers clearer physical interpretations of the results in
quantum cosmology and makes available many  results
obtained previously in the application of the CTP formalism
in quantum field theory in curved spacetimes.

We now proceed to evaluate $\tilde\Gamma$ and $\Theta$  by summing
the equations (\ref{rexp}) and (\ref{phase})
 over all modes $\{k\}$ subject to the restriction  $\a =
\a'$. Some of the mode sums appearing in these expressions are
divergent and hence need to be regularized.
The  regularized influence action  for this problem can be calculated
using Feynman diagram \cite{CH87} or dimensional regularization
techniques \cite{PazSin}.
The phase of the influence functional can be written as
\be
\Theta = {\Gamma}_{div} + {\Gamma}_{ren}
\te
where
${\Gamma}_{div}$ and ${\Gamma}_{ren}$ represent the divergent
and finite contribution to the phase respectively.
${\Gamma}_{div}$ (obtained as terms containing the $1/ \epsilon$ factor in
dimensional regularization , where $\epsilon = n-4$ and $n$ is
the dimension of spacetime ) is given by \cite{PazSin}
\be
{\Gamma}_{div} = \int d{\eta}_1 d {\eta}_2
({\beta}_{ij} - {\beta'}_{ij})({\eta}_1)
{\gamma}_{div} ({\eta}_1 - {\eta}_2)
({\beta}^{ij} + {\beta'}^{ij} )({\eta}_2)
\te
where
\be
{\gamma}_{div}({\eta}_1 - {\eta}_2)
= \int_{-\infty}^{+\infty}{d\omega\over 2\pi} e^{i \omega ({\eta}_1-{\eta}_2)}
\left[ {-{\omega}^4\over 4 {(4\pi)}^2 (n^2 -1)} {1\over \epsilon}\right].
\te
${\Gamma}_{div}$ can be rewritten as
\be
{\Gamma}_{div} = {1\over 4{(4\pi)}^2 (n^2 -1) \epsilon}\int d{\eta}_1
[{\ddot {\beta_i}}^2 -{\ddot {{\beta'}_i}}^2] + surface ~~ terms,
\te
where the surface terms can be written as integrals of total derivatives
of functions of $\beta$ and $\beta'$ and can be discarded.
As it stands this explicitly divergent term cannot be
absorbed by renormalization of the bare coupling constants
present in the original action since from (3.17) we
see that no term of this higher derivative form appears there.
Hence we follow the usual procedure
used in quantum field theory in curved spacetime of first
dimensionally regularizing the effective action, modifiying
the original classical action by adding appropriate counterterms
to cancel the divergence, and finally
taking the limit $\epsilon \rightarrow 0$. The modified classical
action including the counterterms up to second order in $\beta$
is given by \cite{HarHu}
\bea
{\bar S} &=& \int d\eta \left[ -6M {{\dot a}}^2 + {1 \over \{180(4\pi)^2\}}
\left\{ {({\dot a\over a})}^4
   - 3 {({\ddot a\over a})}^2 \right\}\right] \nn \\
 & +& \int d\eta \left(M {\dot \beta}^2 a^2 +  {1 \over \{180(4\pi)^2\}}
\left[ 3 {\epsilon}^{-1}{\ddot \beta}^2
+ 3 ln (\mu a){\ddot \beta}^2 - \left\{ {({\ddot a\over a})} {\dot \beta}^2
 +{({\dot a\over a})}^2 {\dot \beta}^2 - {\ddot \beta}^2 \right\} \right]
 \right)
\tea
where $\mu$ has dimensions of mass and sets the renormalization scale.
The total phase of the density matrix is now given by
\be
{\bar S}( a, \beta) - {\bar S}( a, \beta') + \Theta ( a, \beta, \beta')
\te
Inserting (4.5) for ${\bar S}$ in the above expression we notice
that the pole term in $\epsilon$ cancels exactly. \footnote{However,
we would like to add a cautionary note at this point.
We are assuming without proof that the $R^2$ type terms can be added
as counterterms at this level after making the WKB ansatz.
Since addition of such terms at the level of the quantum cosmology
Hamiltonian which was our starting point involves
the introduction of new canonical degrees of freedom,
the validity of this assumption is not entirely clear.}

The rest of the exponent, ${\Gamma}_{ren}$ and $\tilde{\Gamma}$, is finite:
\be
{\Gamma}_{ren} = \int^{\eta} d{\eta}_1 d {\eta}_2
({\beta}_{ij} - {\beta'}_{ij})({\eta}_1)
{\gamma}_{ren}({\eta}_1 - {\eta}_2)  \label{gren}
({\beta}^{ij} + {\beta'}^{ij} )({\eta}_2)
\te
and
\be
\tilde{\Gamma} = \int^{\eta} d{\eta}_1 d {\eta}_2
({\beta}_{ij} - {\beta'}_{ij})({\eta}_1)
\tilde{\gamma}({\eta}_1 - {\eta}_2)     \label{gtilde}
({\beta}^{ij} - {\beta'}^{ij} )({\eta}_2) ,
\te
where the kernels $\gamma_{ren}$ and $\tilde\gamma$ are given by
\be
\gamma_{ren}(\eta)= -{1\over{60(4\pi)^2}}
\label{odev}
\int_{-\infty}^{+\infty} {{d\omega}\over{2\pi}} ~{\rm e}^{i\omega \eta}
{}~\omega^4 ~
\log(i{{(\omega-i\epsilon)}\over{\mu}})
\te
and
\be
\tilde\gamma(\eta) = ~{1\over{60(4\pi)^2}}
                               \int_{0}^{+\infty} \label{eventilda}
            {{d\omega}\over{2\pi}}{\pi\omega^4} \cos \omega \eta .
\te
Notice that the kernel $\tilde\gamma(\eta)$ is even whereas
$ \gamma_{ren}(\eta)$ contains an odd and even part
given by
\be
\gamma_{odd}(\eta) = ~{1\over{60(4\pi)^2}}
                               \int_{0}^{+\infty} \label{odd}
            {{d\omega}\over{2\pi}}{\pi\omega^4}\sin\omega \eta
\te
and
\be
\gamma_{even}(\eta) =  -{1\over{60(4\pi)^2}}
\int_{-\infty}^{+\infty} {{d\omega}\over{2\pi}} ~\omega^4 \cos
\omega \eta {\rm ln} {|\omega|\over \mu} \label{even}
\te
The kernel $ \gamma_{ren}(\eta)$ is manifestly real and can also be
seen to be causal \cite{CH87}.

Note that $\tilde{\Gamma}$ and ${\Gamma}_{ren}$ play distinct roles here.
$\tilde{\Gamma}$ is responsible for the decoherence between alternative
histories $\beta$ and $\beta'$ in the sense that it suppresses the
contribution of widely differing histories
to the influence functional, and hence  suppresses
the off diagonal terms of  the reduced density matrix .
 This feature and its connection to
particle production was explored before in \cite{PazSin,nfsg}.
On the other hand, when
we attempt to derive the effective equation of motion for $\beta$ by
varying the effective action $S_{eff}$,
only ${\Gamma}_{ren}$ contributes  to generating the equation
of motion. The equation of motion obtained under such variation is
identical to the real, causal dissipative equation for $\beta$ obtained
by Calzetta and Hu in \cite{CH87}. In fact, as we will show more
explicitly later, ${\Gamma}_{ren}$ provides the dissipative
contribution to the equation of motion.
Thus in the present form of the influence functional ${\Gamma}_{ren}$
contributes only to the equation of motion and not to decoherence, and
$\tilde{\Gamma}$ contributes only to decoherence, and not to the
equation of motion.
However, in the following we will show how $\tilde{\Gamma}$ also plays
the dual role of generating noise and will indeed contribute to
the effective equations of motion with a stochastic source.

\subsection{Correspondence with QBM}

Now that we have the complete form of the influence functional,
we can proceed to compare its exponent given by (\ref{gren}) and
(\ref{gtilde})   with that of (\ref{iabm}) of the QBM problem.
We can see that it corresponds to the
$k = 2$,  $f(x) = x$ case in (2.9) with the identification $\beta_i \equiv x$
and $q_n \equiv \chi_{\vk}$. It is by no means obvious that our cosmological
example should correspond to $f(x) = x$, i.e, the linear coupling
case, because in our approximation
we had retained up to quadratic terms in the anisotropy. In fact, from
(\ref{omega}) we see that the system-environment coupling contains
terms quadratic in $\beta$ as well as a quadratic coupling in
velocities, which is not even covered by our Brownian motion model.
However, though these terms are originally present, when correctly
dimensionally regularized, the terms proportional to $\lambda_2$ that
contain the non-linear coupling vanish. Hence we are left with only
an effective linear coupling in the anisotropy.
The local potential renormalization terms $\Delta V$'s can
be identified with $\Gamma_{div}$ in the cosmological case and
we have already dealt with the renormalization.
Using the  time
reflection symmetry of the kernel $\tilde\gamma$ we obtain
\be
\tilde\Gamma=
2~\int\limits_0^\eta d\eta_1\int\limits_0^{\eta_1}d\eta_2
\tilde\gamma(\eta_1-\eta_2) ~({\beta^{ij}} - {{\beta'}^{ij}})(\eta_2)
\label{noise}
\te
and for the phase $\Gamma_{(ren)}$ , using the time reflection
properties of $\gamma_{odd}(\eta)$ and
$ \gamma_{even}(\eta)$ we can rewrite it as
\bea
\Gamma_{ren}& = &
{}~\int^\eta\int^\eta d\eta_1d\eta_2~\beta_{ij}^+(\eta_1)\hat{\gamma}(\eta_1
-\eta_2)\beta^{ij}(\eta_2)\nn \\
& - &~\int^\eta \int^\eta d\eta_1 d\eta_2~{{\beta}'_{ij}}
(\eta_1)\hat{\gamma}(\eta_1 -\eta_2){{\beta}'_{ij}} (\eta_2)\nn \\
& + & 2~\int\limits_0^\eta d\eta_1\int\limits_0^{\eta_1} d\eta_2 ~(\beta_{ij}
-{\beta'}_{ij})(\eta_1) ~ \label{diss}
\gamma_{odd }(\eta_1-\eta_2) ~({\beta^{ij}} + {{\beta'}^{ij}})(\eta_2)
\tea
where
\be
\hat{\gamma}(\eta_1 -\eta_2) = \gamma_{\scriptscriptstyle even}(\eta_1 -
\eta_2)
- \gamma_{\scriptscriptstyle odd }(\eta_1-\eta_2)~sgn(\eta_1 - \eta_2)
\te
is an even kernel. Now we must compare the expressions (\ref{noise})
for the real exponent and the phase (\ref{diss}) with the
corresponding expressions in the influence action (\ref{iabm}) for
the Brownian motion case in order to properly identify the noise
and dissipation contributions. Comparing the real exponents we see
that the noise kernel for the anisotropy in this case is given by
\be
\n(\eta) = 2\tilde{\gamma}(\eta) = ~{1\over{30(4\pi)^2}}
                               \int_{0}^{+\infty} \label{nker}
            {{d\omega}\over{2\pi}}{\pi\omega^4} \cos\omega \eta
\te

In trying to compare the phase terms we notice that the third term in
(\ref{diss}) is indeed of the form of that in (\ref{iabm}) and we can
identify the dissipation kernel $\mu(\eta)$ for the cosmology case as
\be
\mu(\eta) = -2\gamma_{odd}(\eta)                   \label{dker}
\te
and it is manifestly odd in time.

The regularized influence action can therefore be written as
\bea
\Gamma ({\beta},{\beta'}) & = & ~\int^{\eta}\int^{\eta}
d{\eta}_1d{\eta}_2~{\beta_{ij}} ({\eta}_1)\hat \gamma({\eta}_1
-{\eta}_2){\beta^{ij}}({\eta}_2)\nn \\
& - & ~\int^{\eta}\int^{\eta}
d{\eta}_1 d{\eta}_2~{{\beta'}_{ij}} ({\eta}_1)\hat \gamma({\eta}_1
-{\eta}_2){{\beta'}^{ij}}({\eta}_2)\nn \\
& - & ~\int\limits_0^{\eta}d{\eta}_1\int\limits_0^{{\eta}_1} d{\eta}_2
\mu({\eta}_1-{\eta}_2) ~({\beta}^{ij} + {\beta'}^{ij})
(\eta_2)\nn \\
& + & i~\int^\eta d\eta_1\int\limits_0^{\eta_1}d\eta_2
\nu(\eta_1-\eta_2) ~({\beta}^{ij} - {\beta'}^{ij})(\eta_2)
\tea
The first  two terms contribute a non-local potential to the effective
action but  do not contribute to the mixing of $\beta$ and ${\beta'}$
histories like the third and fourth terms. We will now show  in some greater
detail that the
third term with the kernel $\mu$ that is odd in the time domain
contributes to the dissipation and  the last term containing $\nu$
is associated with noise.

\subsection{Noise}

Let us first concentrate on the fourth term. Its contribution to the
 influence functional is given by
\be
exp [-~\int^\eta d\eta_1\int\limits_0^{\eta_1}d\eta_2
\nu(\eta_1-\eta_2) ~({\beta}^{ij} - {\beta'}^{ij})
(\eta_2)] \label{expnoise2}
\te
We will proceed in exact analogy with the analysis of noise in the
case of QBM described in Sec. 2.2.
The term in (4.19) can be rewritten using  functional Gaussian identity
(2.11)
which in this case states that the above expression  is equal to
\be
\int  D\xi(\eta) {\cal P}[\xi]exp [ i\int\limits_0^\eta
d{\eta'} {\xi}^{ij}({\eta'})~ ({\beta}_{ij} - {{\beta'}_{ij}})({\eta'})]
\te
where
\be
{\cal P}[\xi] = P_0 exp[{-                                \label{noisedist}
}\int\limits_0^\eta d{\eta}_1\int\limits_0^{\eta}d{\eta}_2
\ha {\xi}_{ij}(\eta_1)\nu^{-1}
(\eta_1 - \eta_2){\xi}^{ij}(\eta_2)]
\te
is the functional distribution of $\xi(\eta)$ and $P_0$ is a
normalization factor given by
\be
{P_0}^{-1} = \int D\xi(\eta)exp [- \int\limits_0^\eta d\eta_1
\int\limits_0^\eta d\eta_2{\xi}_{ij}(\eta_1) \nu^{-1}(\eta_1-\eta_2)
{\xi}^{ij}(\eta_2)].
\te
The influence functional can then be written as
\bea
e^{i\Gamma} &=& \int D\xi(\eta){\cal P}[\xi] exp{i\hat{\Gamma}
[\beta, \beta' ,\xi]}\nn \\
&\equiv & {< exp {i\hat{\Gamma}
[\beta, \beta' ,\xi]}  >}_{\xi}
\tea
where the angled brackets denote an average with respect to the
stochastic distribution ${\cal P}[\xi]$.
The modified influence action $\hat{\Gamma}[ \beta, \beta', \xi]$
is given by
\bea
\hat{\Gamma}[ \beta , \beta', \xi] & = &
  ~\int^{\eta}\int^{\eta}
d{\eta}_1d{\eta}_2~{\beta}_{ij} ({\eta}_1){\hat \gamma}({\eta}_1
-{\eta}_2){\beta^{ij}}({\eta}_2)\nn \\
& - & ~\int^{\eta}\int^{\eta}
d{\eta}_1 d{\eta}_2~{\beta'}_{ij} ({\eta}_1){\hat \gamma}({\eta}_1
-{\eta}_2){{\beta'}^{ij}}({\eta}_2)\nn \\
& - & ~\int\limits_0^{\eta}d{\eta}_1\int\limits_0^{{\eta}_1} d{\eta}_2
\mu({\eta}_1-{\eta}_2) ~({\beta}^{ij} + {\beta'}^{ij})
(\eta_2)\nn \\
&-& ~\int d\eta' {\xi}^{ij}(\eta'){\beta}_{ij}
   +  ~\int d\eta' {\xi}^{ij}(\eta'){\beta'}_{ij}
\tea
The  term coupling a stochastic source $\xi$ to $\beta$ will manifest itself as
the noise in the equation of motion derived from this effective action. We see
that the influence action $\Gamma$ can be  written as an average of
$\hat{\Gamma}$ over this stochastic distribution function.

The reduced density matrix  can thus also be written as a stochastic average
\be
{\rho}_{red}[ {\beta},{\beta'}]
 =  <e^{i \hat S_{eff} (
{\beta},
{\beta'};\xi)}>_\xi
\te
where the full effective action $\hat S_{eff}$ is given by
\bea
\hat S_{eff} &=&
{\bar S}[a, {\beta}] + ~\int d\eta' {\xi}^{ij}(\eta'){\beta}_{ij}
- \{{\bar S}[a , {\beta'}] + ~\int d\eta' {\xi}^{ij}(\eta'){{\beta'}_{ij}}\}
\nn \\
& - & ~\int\limits_0^{\eta}d{\eta}_1\int\limits_0^{{\eta}_1} d{\eta}_2
\mu({\eta}_1-{\eta}_2) ~({\beta}^{ij} + {\beta'}^{ij})
(\eta_2) \label{seff}
\tea
and $\bar S$ is given by  (4.5).
Our relevant equations of motion will be derived by varying $\hat S_{eff}$.
{}From  this equation we can view ${\xi}(\eta)$ as an external stochastic
force linearly coupled to $\beta$, though the linearity is a feature
specific to truncation of the perturbation series at quadratic order in the
effective action. In general we will have non-linear coupling.

Since the distribution functional (\ref{noisedist}) is Gaussian, this is a
Gaussian type noise, which as in (2.18) , is completely characterized by
\bea
{<\xi(\eta)>}_{\xi} &=& 0 \nn\\
{<\xi(\eta_1)\xi(\eta_2)>}_\xi &=& \nu(\eta_1-\eta_2)
\tea
Therefore the non-local kernel $\nu(\eta_1-\eta_2)$ is just the two-point
time-correlation function of the external stochastic source
$\xi(\eta)$. Since this correlation function is non-local, this noise is
colored.  As suggested in \cite{HuPhysica,HMLA}
we believe this is a rather general feature of noise of cosmological origin.

\subsection{Fluctuation-Dissipation Relation}

Now that we have identified the noise and dissipation kernels
$\nu(\eta)$ and $\gamma_{odd}(\eta)$ respectively, we can
go ahead and write down the fluctuation-dissipation relation in
analogy  with the quantum Brownian model \cite{HPZ2,HuBelgium}.
Defining
\be
\mu(\eta) = - 2 \gamma_{odd}(\eta) = {d\over d\eta}\gamma(\eta)
\te
The fluctuation-dissipation relation has the familiar form
given by (2.30)
\be
\nu(\eta) = \int\limits_0^\infty d{\eta}' K(\eta-{\eta}')\gamma({\eta}')
\label{fdan}
\te
where the FD kernel $K(\eta)$ is given by
\be
K(\eta) = \int\limits_0^{\infty} {d\omega\over \pi}~\omega \cos\omega \eta
\te
This  supports the conjecture   of \cite{HuPhysica} that there exists
a fluctuation-dissipation relation for the description of the backreaction
effect of particle creation in cosmological spacetimes.
We see that the FD kernel is identical with that given by  (\ref{zerofd}),
which is given for more general system-bath couplings of the form (\ref{int}),
but with the bath at $T=0$.  Hence this also vindicates
the previous observation  \cite{HPZ2,Zhang,SinSor} that the zero temperature
fluctuation-dissipation relation is insensitive  to the nature of the
system-bath coupling. Since we have not taken the bath at a finite
temperature, thermal fluctuations play no role in the above relation
and it summarizes the effect solely of quantum fluctuations. Effect of
thermal fluctuations can be included easily and we expect a FDR to hold
for finite temperature particle creation and backreaction as well.

\section{Particle Creation, Noise and Backreaction}

\subsection{Particle Creation}

In this section we would like to examine in some detail the
relationship between the noise and dissipation kernels and particle
creation from the vacuum.
We would also be interested in comparing this approach to
that in \cite{CH87} and \cite{HuPhysica} where the relationship between
particle production and anisotropy dissipation was discussed in some
depth.

Let us first concentrate on the noise term . Since we know that the
noise term comes from the real part $\tilde{\Gamma}$
of the exponent of the influence
functional, we will analyze  this part and try to rewrite in a form
such that it is easy to identify the part associated with particle
production. It can be shown \cite{PazSin} that $\tilde{\Gamma}(\beta,{\beta'})$
can be rewritten as

\be
\tilde\Gamma(\beta,{\beta'}) =\int {d^3k \over {(2\pi)}^3}{1\over{4\ok^2}}
|B_k(\beta) - B_k({\beta'}) |^2
\te
where
\be
B_k(\beta) = -i {{\ok}\over 2}~
\int^\eta d\eta_1 ~{1\over\ok} ~[2 {\beta}_{ij}(\eta_1)k^i k^j]
\te
and $\ok = |{\vec k}| = {(\sum_i {k_i}^2)}^{1\over 2}$. As we may recall
from (3.33), the term $2 {\beta}_{ij}(\eta_1)k^i k^j = \lambda_1$ is the
expansion
of the natural frequency to the first anisotropy order. One can of course
go to higher orders.

Now we can show that a close relation exists between the $B(\beta)$
function and the Bogolubov coefficients associated with the particle
creation that takes place as a consequence of the anisotropy
evolution \cite{HarHu,HuPar78}. This can be seen as follows.

The conformally related massless scalar field $X = a\Phi$ in our
model can be decomposed into modes as
\be
X = \int {d^3k \over {(2\pi)}^3} e^{i {\vec k}\cdot{\vec x}} \chi_k(\eta)
\te
$\chi_k$ satisfies the following equation to first order in anisotropy
\be
{d^2 \chi_k \over d{\eta}^2} + ( {\ok}^2 + 2 {\beta}_{ij}(\eta_1)k^i k^j)
\chi_k  = 0  \label{waveq}
\te
The solution to the above equation ( again to first order in $\beta$)
is given by
\bea
\chi_k(\eta) & = & {\chi_k}^{in}(\eta)~\Bigl[1+ \int^{\eta} d\eta_1 \nn \\
     & &  - {{\chi_k}^{in}}^* (\eta) \int^\eta d\eta_1 ~{1\over{2i\ok}}
      [2 {\beta}_{ij}(\eta_1) k^i k^j]   ~e^{2i\ok \eta_1}      \label{soln}
\tea
where
\be
{\chi_k}^{in}(\eta)  = {1\over{{\sqrt{2\ok}}}} e^{i\ok \eta}
\te
is the solution to (\ref{waveq})
with $\beta_{ij} = 0$ and corresponds to the `in'
conformal vacuum in the far past. Assuming that the anisotropy is switched
off at time $\eta$ the term on the left hand side of (\ref{soln}) can be
associated with ${\chi_k}^{out}(\eta)  $  the `out' vacuum.
As is well known, the ``in" and ``out" basis can
be related in terms of Bogolubov coefficients $\alpha_k$ and ${\hat{\beta}_k}$
as
\be
{\chi_k}^{out}(\eta) =  \alpha_k {\chi_k}^{in}(\eta) + {\hat{\beta}}_k
{{\chi_k}^{in}}^* (\eta)
\te
Comparing (5.5 ) and (5.7 )  we can identify the ${\hat{\beta}}_k$ Bogolubov
coefficient as
\be
{\hat{\beta}}_k =
\int^\eta d\eta_1 ~{1\over{2i\ok}}( 2 {\beta}_{ij}(\eta_1) k^i k^j)
\te

 As we see from its definition, the function
$B$ is proportional to this Bogolubov coefficient $\hat{\beta}$.

\be
B_k(\beta) = \ok e^{-2i\ok \eta} \hat{\beta}_k
\te
Thus
\be
\tilde\Gamma(\beta,{\beta'}) =\int {d^3k \over {(2\pi)}^3}{1\over 4}
|{{\hat{\beta}}_k} - {{\hat{\beta}}_k}'  |^2,
\te
where $ {{\hat{\beta}}_k} $ and $ {{\hat{\beta'}}_k}$ are the
Bogolubov coefficients associated with the anisotropy histories $\bp$ and
$\bm$ respectively. It is obvious from (5.19 ) that the noise will be
non-zero only provided   $ {{\hat{\beta}}_k}  \neq {{\hat{\beta'}}_k} $,
i.e, if there is different amounts of particle production along the
two histories. Since this term is also associated with decoherence this
is also a necessary condition for decoherence to occur. This has also
been noticed from a slightly different point of view in \cite{nfsg,CalMaz}.

This demonstrates a connection between the process of particle production
and the noise or fluctuation.

\subsection{Einstein-Langevin Equation}

We will now show how this noise can be incorporated into the equation of
motion as a Langevin type equation. In this process we will also
demonstrate the role of the kernel $\mu$ in providing dissipation. The
key difference from the earlier treatment \cite{CH87} is that
the equation of motion will be derived from  the quantity
$\hat S_{eff}( \beta, \beta', \xi)$ rather than the ``noise
averaged" quantity ${S_{eff}(\beta, \beta')}$.
This has also been discussed in other contexts in \cite{HPZ2,nfsg,HM3}.
The first step is to write $\hat S_{eff}( \beta , \beta' , \xi)$
in terms of the following variables
\bea
\bar\beta_{ij} &=& {1\over 2} ( \bp + \bm) \nn \\
\Delta             &=& \bp - \bm
\tea
The equation of motion is then derived as
\be
{\delta \hat S_{eff}( \bar \beta_{ij} , \Delta)\over \delta \Delta }
{\Big | }_{\Delta = 0} = 0
\te
yielding
\bea
& & -2 M {d\over d\eta} ( a^2 {\dot {\bb}}) + {1\over 30{(4\pi)}^2}
{d^2\over d{\eta}^2} [{\ddot {\bb}}ln(\tilde \mu a)]
+ {1\over 90{(4\pi)}^2} {d\over d\eta}
\left\{ \left[ {\Big({{\dot a}\over a}\Big)}^2
+ \Big({{\ddot a}\over a}\Big)\right]
{\ddot {\bb}}\right\}\nn \\
& + & \int d{\eta}_1 \gamma_{ren} (\eta - \eta_1)\bb(\eta_1)
= - j_{ij}(\eta) + {\xi}_{ij}(\eta)             \label{eom}
\tea
Here $j_{ij}$ is an external source term added in order to switch on
the anisotropy in the distant past \cite{HarHu}.
It is worth comparing these results with those in \cite{CH87} where
similar equations were deduced from the CTP effective action.
Comparing (\ref{eom}) with (3.18) in \cite{CH87} we find
that they are exactly the same except for the stochastic force ${\xi}_{ij}$
on the right hand side.
The real and causal kernel $K_4$ there (including the
numerical factor $1/[30(4\pi)]^2$) is identical to our kernel
$\gamma_{ren}$ . We will show that the odd part of this
kernel can be associated with dissipation.
One could  in fact interpret (3.18) obtained by Calzetta and
Hu as (\ref{eom}) averaged with respect to the noise
distribution. Since this is a Gaussian noise, $<\xi> = 0 $,
we obtain (3.18) of \cite{CH87}, where the
$\beta$'s are also to be interpreted as noise-averaged variables.
In this sense, we have gone beyond previous analysis
in  extracting the underlying stochastic behavior
that is lost in the smoothed out average version given in \cite{CH87}.

To make the analogy with a Langevin equation more explicit it is convenient
to integrate (\ref{eom}) once with respect to $\eta$. This gives the following
equation

\bea
& & -2 M a^2 {\dot {\bb}} + {1\over 30{(4\pi)}^2}
{d\over d{\eta}} [{\ddot {\bb}}ln(\tilde \mu a)]  + {1\over 90{(4\pi)}^2}
\left\{ \left[ {\Big({{\dot a}\over a}\Big)}^2
+ \Big({{\ddot a}\over a}\Big)\right]
{\ddot {\bb}} \right\}\nn \\
& + & \int d{\eta}_2\int d{\eta}_1 \gamma_{ren} (\eta_2 - \eta_1)\bb(\eta_1)
 = c_{ij} + s_{ij}
\tea
where $c_{ij}(\eta) = -\int d{\eta}' j_{ij}({\eta}')$ and
$s_{ij}(\eta) = \int d{\eta}' {\xi}_{ij}({\eta}')$.

Defining the variable $q_{ij} = d{\bb}/ d\eta$ we can write the above
equation  in the following  form
\be
{d\over d\eta}(\tilde M {dq_{ij}\over d\eta}) + {\cal K} {dq_{ij}\over d\eta}
+ k q_{ij} = c_{ij} + s_{ij}      \label{1int}
\te
where
\bea
\tilde M &=& {1\over 30{(4\pi)}^2}ln(\tilde \mu a) \\
k &=&  -2 M a^2   + {1\over 90{(4\pi)}^2}
\left[ {\Big({{\dot a}\over a}\Big)}^2 + \Big({{\ddot a}\over a}\Big)\right] \\
{\cal K}q_{ij} &=& \int d{\eta}_2\int d{\eta}_1 f(\eta_2 - \eta_1)
                   {dq_{ij}\over {d\eta_1}}
\tea
and $ d^{2}f(\eta)/ d{\eta}^2 = \gamma_{ren}$. This equation
is identical in form to the equation (3.15) in \cite{HuPhysica}
except for the term $s_{ij}$ on the right hand side, which is
indeed the stochastic contribution  from the noise anticipated there.
This equation is a generalized  Einstein equation in the Langevin form,
in that there is a dissipative term in the dynamics and a noise term in
the source. It has been conjectured \cite{HuPhysica}
and shown \cite{nfsg} that in a more complete description of semiclassical
gravity the semiclassical Einstein equation driven by the expectation values of
the energy-momentum tensor should be replaced by an Einstein-Langevin equation,
where there is an additional stochastic source arising from the fluctuations
of quantum fields. The conventional semiclassical Einstein equation is in this
sense, the simplified mean-field theory.

\subsection{Dissipation and Backreaction}

Equation (\ref{1int})  is in the form of a generalized
damped harmonic oscillator driven by a stochastic force $s_{ij}$.
(Of course the generalized mass $\tilde M$ and spring constant $k$ are time
dependent, so strictly speaking it has the damped harmonic oscillator
analogy only when these quantities are positive, as was also pointed
out in \cite{Paz90}.)

The second term on the left hand side of (\ref{1int})
represents the damping term involving a non-local (velocity dependent)
friction force. That this term is associated with dissipation can
be quickly seen as follows \cite{CH89}. In the Fourier transformed version
of a damped harmonic  oscillator equation the imaginary term
is associated with dissipation. Writing  (\ref{1int}) in terms of the
Fourier transform $q_{ij}(\omega) = \int d\eta e^{-i\omega \eta} q_{ij}(\eta)$
we notice that the only imaginary contribution comes from the
second term on the left hand side, which can be written as
\be
F(q) = \int{d\omega\over 2\pi} e^{i\omega\eta}
{\gamma_{ren}(\omega)\over {\omega}^2} q_{ij}(\omega)
\te
where $\gamma_{ren}(\omega)$ is the Fourier transform of $\gamma_{ren}(\eta)$
defined in  (\ref{odev}). Thus we see that the dissipation is associated
with the imaginary part of $\gamma_{ren}(\omega)$ or equivalently with
the odd part of the kernel $\gamma_{ren}(\eta)$ given by
$\gamma_{odd}$ defined in (\ref{odd}). This is consistent with our earlier
identification of $\gamma_{odd}$
as the dissipation kernel from the form of the influence functional
compared with the Brownian motion case.
In fact, as in \cite{CH87,CH89} we can isolate the generalized (frequency
dependent) viscosity function $\zeta(\omega)$ by writing
\be
i\zeta(\omega)\omega q_{ij}(\omega) =
i {\rm Im } \gamma_{ren}(\omega)q_{ij}(\omega)
\te
{}From  (\ref{odev}) we can identify $\zeta(\omega)$ as
\be
\zeta(\omega) = {{|\omega|}^3\over 60(4 {\pi})^2}
\te
which is identical to that found in \cite{HuPhysica} and which is not
surprising, since our kernel $\gamma_{(ren)}$ in (\ref{eom}) and
$K_4$ in  (3.18) in \cite{CH87} are identical up to numerical
factors.

Once having made the identification of the velocity dependent viscous
force in the equation of motion , we can calculate the dissipated
energy density by integrating $\vec{F}.\vec{v}$ (with $\vec v =\dot q_{ij}$
acting as velocity) over all frequencies
and come up with an expression identical to (3.18) in \cite{CH87,CH89}.
\be
\rho_{dissipation} = \int\limits_0^{\infty}{d\omega\over 2\pi}
[\omega{\beta_{ij} (\omega)}^*][\zeta(\omega)\omega\beta_{ij}(\omega)].
\te
which has been shown there to be identical to the total energy of particle
pairs created by a given anisotropy history $\beta$. In this way, we
can see the connection between particle production and the
dissipation kernel and hence the process of dissipation itself.
Earlier in this section we had demonstrated the connection between
particle production and the noise or fluctuation term. On the other
hand, (\ref{fdan}), the fluctuation-dissipation relation, embodies a
relationship between the processes of fluctuation and dissipation of
anisotropy. So this completes the full circle of connections among
these processes. In a way, one can say that as a physical process,
particle production is contributing to both the noise and
dissipation, and of course these are two different manifestations of
the loss of information due to integrating over the field modes.

\section{Discussion}

In closing, we would like to discuss the meaning of the FDR in semiclassical
cosmology in a broader context and mention some related problems
for future investigation.\\

1) {\it FDR under Finite Temperature and Non-Equilibrium Conditions}

In this paper we have discussed in detail the FDR in semiclassical cosmology
under a zero temperature bath. A similar relation between the noise and
dissipation kernels exists for baths at finite temperature. The form will
be similar to that derived for the QBM problem in Sec. 2 \cite{HPZ2}.
One can take the
finite temperature calculation via the CTP formalism \cite{Paz90} and
perform a similar analysis as we have done for the vacuum case and obtain the
results explicitly. In reality both
vacuum and thermal bath results will enter into the picture,\footnote{As
has been discussed earlier \cite{ftf}, the energy density of the quantum
field at any moment will contain two parts. There is a zero temperature
component and a finite temperature component, the former corresponds to
spontaneous creation from the vacuum, and the latter is of the nature of
stimulated creation from particles already present
\cite{Par69,HuPav,HuKan,HKM}.}
since once particle creation commences,  given sufficient time and assuming
some
(collisional) interaction amongst the created particles,
the bath will soon acquire a finite temperature character.\footnote{A finite
temperature bath at every moment is only an idealization.
To use a finite temperature description one has to discriminate the conditions
for the bath to thermalize, and for the system to be equilibrated with it.
These vary with the nature of the bath (massive or massless,
linear or nonlinear interactions, spectral density) and the form of interaction
between the system and the bath. See the analysis of \cite{HPZ3,Gle,Boy}}
This heat-up process is expected to happen quickly near the Planck time,
especially so for anisotropic universes, as particles are created profusely
there \cite{ZelSta,Hu74}, generating a large amount of entropy
\cite{Hu81,Hu84,HuPav,HuKan,BMP,GV}.
The copious creation of particles near the Planck time is accompanied by large
fluctuations and noise, and it induces a strong backreaction on the spacetime
dynamics, dissipating the anisotropy rapidly \cite{ZelSta,HuPar78,HarHu}.
The weaker anisotropy in the universe's expansion induces lesser particle
creation. The lower particle creation rate is accompanied
by a smaller fluctuation and noise, which in turn gives weaker dissipation
of spacetime anisotropy.
The surviving anisotropy would continue to sustain particle creation,
albeit in much smaller amounts. And this goes on.
(The backreaction  follows a Lenz law behavior which was expounded
in earlier studies \cite{Par69,Hu83,Hu84}.)
At each stage we expect to see a balance between the rate of particle creation
and the strength of fluctuations and dissipation.

In reality the spacetime-field combined system involving
particle creation exists in a highly non-equilibrium state.
To give a quantitative description of the above processes one needs to
describe the dynamics of the actual statistical state of both the system and
the  environment in a self-consistent manner, which is a highly non-trivial
problem. What we have described in this paper is only the first step, which
depicts the effect of particle production from a vacuum (zero-temperature)
bath.
The interaction of created particles and how they alter the environment
(e.g., thermalization) is not accounted for.
In the second step one needs to also examine the evolution of
the environment (quantum field) taking into consideration the effects of
spontaneous and induced particle creation, their interaction and the entropy
generation processes, all in the context of a changing background spacetime
whose dynamics at each moment affects and is also affected by
the activities of this environment.

Despite all these complexities, even in highly nonequilibrium conditions we
expect that a generalized FDR (in the form given in this paper)
will still hold and be useful to guide us on understanding the complex
physical processes in the system and the environment. From the above
depiction of the physical scenario and from previous studies of
the statistical mechanics of quantum field processes in cosmology,
one can see that there exists a balance between particle creation (in the
field) and its backreaction (on spacetime), which can be attributed to the
interlocked relation between fluctuations and dissipation. There is also
a mathematical justification: it is a relation between the real and
imaginary parts of the effective action for the open system.
Similar in nature to the optical theorem in scattering theory
or the Kramers-Kr\"oning relation in  many body theory \cite{nfsg},
these relations describe the dissipative and
reactive parts of the response function of an open system to influences
from the environment. They are of a categorical nature because
they originate ultimately from the unitarity condition of the dynamics of
the combined closed system. They only take the form of
dissipation in the open system because we have identified a certain subsystem
as the system of interest and decided to follow its effective dyanamics;
and they take the form of fluctuations in the environment because we
refer to them in reference to the mean value of the environment variables,
the remaining information is downgraded in the form of fluctuations.
Had we decided not to coarse-grain the environment, or choose
to observe the two subsystems with equal interest and accuracy,
such a relation governing the mutual reaction would still exist,
except that the concepts of dissipation and  fluctuations will no longer
be appropriate.
(Both subsystems will be governed by equations of motion in the form
of an integral differential equation, and treated in a nondiscriminate
and balanced way. See, e.g., \cite{ProjOp}).

In the context of semiclassical
cosmology, the open system is the spacetime sector,
whose dynamics is influenced by the matter fields.
The expansion of the universe amplifies the vacuum fluctuations
of the matter field into particles, which act as the source in the Einstein
equation driving the universe. The averaged effect of particles
created imparts a  dissipative component in the spacetime dynamics,
and the fluctuations in particle creation constitute the noise.
The particular forms of the dissipation and noise kernels and their effects
may vary under different conditions-- zero or finite temperature,
equilibrium or non-equilibrium-- but the existence of such a relation
between the fluctuations in the matter field and
the dissipative effect on the spacetime dynamics should remain.
We will have opportunities later to explore related problems which can shed
more light on these issues.\\

2) {\it Relation with FDR in Spacetimes with Event Horizons}

As we mentioned in the Introduction, our search for a FDR in cosmological
spacetimes without event horizons was inspired by Sciama's proposal to
view the Hawking and Unruh effects as manifestations of a fluctuation-
dissipation relation between the field quanta and the detector response.
De Sitter universe is an important class of  cosmological spacetimes
with event horizons. For this one can use the thermal property of the
field to perform a linear response theory (LRT) analysis for the derivation of
the FDR \cite{Mottola}. Our derivation here based on
the influence functional formalism attacks the problem at a more basic level,
where equilibrium condition between the system and the environment is
not necessarily present at every stage.
It is of interest to compare the results between
the equilibrium limit of the IF or the CTP formalisms and that of LRT.
This can be done explicitly by carrying out an analysis similar to
this paper on the de Sitter universe and see how the FDR obtained from
the IF compare with that from the LRT. Formally this would render
explicit the relation between the IF formalism to (non-equilibrium)
statistical field theory and perturbative thermal (finite-temperature)
field theory.

More meaningfully, as was
originally concieved by one of us \cite{HuEdmonton,HuPhysica}, this would
provide a channel to generalize the conventional way of treating
Hawking effect associated with black holes and accelerated observers
based on thermal propagators and event horizons to non-stationary conditions.
This involves cases like non-uniformly accelerating observers and
realistic collapse dynamics, where an event horizon or Euclidean section
does not always exist but is dynamically generated.
Our motivation for finding a way to treat these
more general conditions is to seek a deeper meaning to the Hawking effect,
and through it to explore the subtle connection between quantum field theory,
relativity theory and statistical mechanics.
In our view, the open system concept explicated by the influence functional
formalism provides a more solid basis to understand its statistical
mechanical meanings and a broader framework to tackle the less unique
situations which cannot easily be treated by purely geometric means,
powerful and elegant as they are.
It also brings the effects of quantum fields on observer kinematics
and spacetime dynamics in line with the more common  statistical mechanical
phenomena involving ordinary matter.
These problems are currently under investigation.\\

3) {\it Related Problems in Semiclassical Cosmology and Inflationary Universe}

For particle creation-backreaction problems similar to the Bianchi-I model
studied here, one can obtain similar results for other matter fields in
other types of spacetimes of astrophysical or cosmological interest.
An example is a massless minimally-coupled scalar field in a
Robertson-Walker or de Sitter universe.
It mimics the linearized graviton modes and has practical use for the
description of primordial stochastic gravitons. The particle production
problem was first studied by Grishchuk \cite{Gri}, the backreaction by
Grishchuk \cite{Gri76} and  Hu and Parker \cite{HuPar77} via canonical
quantization methods, and by Hartle \cite{Har81}
and Calzetta and Hu \cite{CH87} via the in-out and in-in effective action
method respectively. The influence functional approach
expounded here would enable one to get from first principle the entropy
generation from graviton production \cite{BMP,GV},
and the noise associated with them, which is related to
the fluctuations in graviton number \cite{HKM,nfsg},
On the aspect of backreaction in semiclassical cosmology,
one can also derive the Einstein-Langevin
equation for the study of graviton production and metric fluctuations.
This problem is currently pursued by Calzetta and Hu \cite{CHfluc}

A related problem of interest is the evolution of the homogeneous mode of
the inflaton which describes the inflation mechanism \cite{decinf}
and the inhomogeneous modes as progenator of structures in the early universe.
The influence functional method was used by Hu, Paz and Zhang \cite{HuBelgium}
Laflamme and Matacz \cite{LafMat} and others to discuss the
decoherence of the long-wavelength sectors of the inflaton, and
the origin of quantum fluctuations as noise for the galaxy formation problem.
Our result here provides an example for the consistent treatment of
the evolution of these modes, their intereaction, and their backreaction
on the spacetime, which can offer some  physical insight into the
no-hair type of theorems in inflationary universe. These problems are
under study by Matacz, Raval and the authors.\\

4) {\it Minisuperspace in Quantum Cosmology as an Open System:
        Geometrodynamic Noise and Gravitational Entropy}

We have discussed the question of the validity of
the minisuperspace approximation \cite{mss} in quantum cosmology
\cite{SinHu,Sinha,HPS}, wherein only the homogeneous cosmologies
are quantized and the inhomogeneous cosmologies ignored \cite{HuErice}.
We used an interacting quantum field model and calculated the effect of
the inhomogeneous modes on the homogeneous mode via the CTP effective action.
This effect manifests in the effective equation of motion for the system as a
dissipative term. For quantum cosmology, this backreaction
turns the Wheeler-De Witt \cite{WdW}
equation for the full superspace into an effective
equation for the minisuperspace with dissipation.
Extending the CTP to the IF formalism as is done here, one can derive
the noise associated with the truncated inhomogeneous cosmological modes.
One can also define an entropy function from the reduced density matrices,
which measures the information loss in the minisuperspace truncation.
These can perhaps be called geometrodynamic noise and gravitational entropy.
It would be interesting to compare this statistical mechanical
definition with the definition suggested by Penrose \cite{Penrose} in classical
general relativity and by one of us in the semiclassical context \cite{Hu83}.
Some initial thoughts on this problem are described in \cite{HuWaseda}, while
details are to be found in \cite{HuSinSTN}.\\

{\bf Acknowledgements} We thank Esteban Calzetta and Juan Pablo Paz
for interesting discussions. Research is supported in part by the
National Science Foundation under grant PHY91-19726.


\end{document}